\documentclass[lengthestimate,tighten,amssymb,prb,aps,twocolumn,byrevtex,floatfix,tightenlines,superscriptaddress]{revtex4}

\usepackage{epsfig}

\newif\ifpdf
	\ifx\pdfoutput\undefined
	\pdffalse 
	\else
	\pdfoutput=1 
	\pdftrue
\fi

\ifpdf
	\usepackage[pdftex]{graphicx}
	\else
\usepackage{graphicx}
\fi

\begin{document}
\ifpdf
	\DeclareGraphicsExtensions{.pdf, .jpg, .tif}
	\else
	\DeclareGraphicsExtensions{.eps, .jpg}
\fi

\title{Demonstration of an all-optical quantum controlled-NOT gate}
\author{J. L. O'Brien}
\affiliation{These authors contributed equally to this work}
\affiliation{Centre for Quantum Computer Technology, Department of Physics, University of Queensland 4072, Australia}
\author{G. J. Pryde}
\affiliation{These authors contributed equally to this work}
\affiliation{Centre for Quantum Computer Technology, Department of Physics, University of Queensland 4072, Australia}
\author{A. G. White}
\author{T. C. Ralph}
\affiliation{Centre for Quantum Computer Technology, Department of Physics, University of Queensland 4072, Australia}
\author{D. Branning}
\affiliation{Centre for Quantum Computer Technology, Department of Physics, University of Queensland 4072, Australia}
\affiliation{Department of Physics, University of Illinois at Urbana-Champaign, Urbana Illinois 61801-3080, USA}
\date{\today} 

\maketitle

\textbf{
The promise of tremendous computational power, coupled with the development of robust error-correcting schemes\cite{nielsen}, has fuelled extensive efforts\cite{qic-special-issue} to build a quantum computer. The requirements for realizing such a device are confounding: scalable quantum bits (two-level quantum systems, or qubits) that can be well isolated from the environment, but also initialized, measured and made to undergo controllable interactions to implement a universal set of quantum logic gates\cite{di-sm-23-419}. The usual set consists of single qubit rotations and a controlled-NOT (CNOT) gate, which flips the state of a target qubit conditional on the control qubit being in the state 1. Here we report an unambiguous experimental demonstration and comprehensive characterization of quantum CNOT operation in an optical system. We produce all four entangled Bell states as a function of only the input qubits' logical values, for a single operating condition of the gate. The gate is probabilistic (the qubits are destroyed upon failure), but with the addition of linear optical quantum non-demolition measurements, it is equivalent to the CNOT gate required for scalable all-optical quantum computation\cite{kn-nat-409-46}.}

Nuclear magnetic resonance techniques have been used to implement the most advanced quantum algorithms to date\cite{va-nat-414-883}. As they encode in mixed states and make ensemble measurements, these systems are not ultimately scalable. In contrast, ion trap systems have been used to implement high-fidelity two qubit quantum gates on pure states of trapped ions\cite{sc-nat-422-408,le-nat-422-412}. Solid-state systems, including spin qubits in semiconductors\cite{ka-nat-393-133} and quantum dots\cite{qic-special-issue}, have been hailed for their potential scalability, and recently, entanglement between two superconducting qubits was demonstrated\cite{pa-nat-421-823}. Single photon qubits offer the dual advantages of excellent isolation from the environment and ease of manipulation at the single qubit level, and have consequently found wide application in quantum cryptography protocols\cite{gi-rmp-74-145}. Photon qubits have the added advantage that an unequalled level of control over their quantum state makes comprehensive characterization possible, as demonstrated in the state tomography measurements presented here.

\begin{figure}[t!]
\begin{center}
\includegraphics*[width=7.5cm]{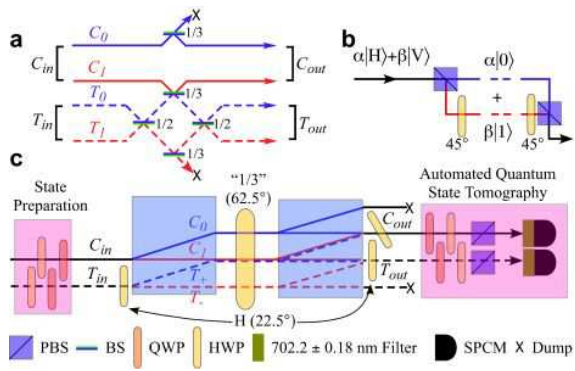}
\caption{A schematic of the CNOT gate realized in this work. (a) A conceptual depiction of the gate, as described in the text. A sign change ($\pi$ phase shift) occurs upon reflection off the green side of the BSs (b) A polarisation encoded photonic qubit can be converted into a spatially encoded qubit, suitable for the gate shown in (a), using a PBS and a HWP set to rotate the polarisation of one of the outputs by 90$^\circ$ (OA = 45$^\circ$). The rotation is required so that all components of the spatial qubits have the same polarisation and can interfere both classically and non-classically. The reverse process converts the spatial encoding back to polarisation encoding. (c) A schematic of the experimental CNOT gate. Pairs of energy degenerate photons are incident from the left of the diagram. These were generated through beam-like spontaneous parametric downconversion \protect\cite{ta-ol-26-843} and collected into single mode optical fibres\protect\cite{ku-pra-64-023802} (not shown). The output of each fiber is collimated and a HWP and QWP in each input beam allows preparation of any pure, separable two qubit state to be input into the gate. The horizontal and vertical components of the qubits are separated and recombined using PBS made from the birefringent material calcite, where the output modes are parallel and displaced. This interferometer is inherently stable, being insensitive to translation of the PBSs. The two outputs are polarisation analysed using an automated tomography system consisting of a computer controlled HWP and QWP followed by a PBS in front of each single photon counting module (SPCM). Simultaneous detection of a single photon at each of the detectors---a coincidence count---signals that the gate has worked. A coincidence window of 5 ns was used throughout. The tilted HWP at 0$^\circ$ is fixed to correct a phase shift in the control interferometer.}
\label{schematic}
\end{center}
\end{figure}

The difficulty in optical quantum computing has been in achieving the two photon interactions required for a two qubit gate (although progress has been made in cavity quantum electrodynamics systems\cite{tu-prl-75-4710}). Knill, Laflamme and Milburn (KLM) have proposed a solution: a non-deterministic CNOT gate, where the required nonlinearity is accomplished using extra 'ancilla' photons---photons that are not part of the computation---and single photon detection. This gate can be made efficiently deterministic (scalable) by a teleportation protocol\cite{go-nat-402-390}. Related proposals require triggered entangled pairs of photons as a resource\cite{ko-pra-63-030301,pi-pra-64-062311}. Other two-photon gates show some, but not all, of the features of a quantum CNOT gate\cite{pa-nat-410-1067,pi-prl-88-257902} in that they cannot work for an arbitrary input state. The ultimate realisation of the KLM CNOT gate will require: heralded single photon sources with stringent mode and bandwidth characteristics; high-efficiency number resolving single photon detectors; and construction of complicated optical circuits exhibiting both classical and quantum interference effects. Progress has been made towards reaching the first two requirements\cite{sa-nat-419-594,ku-prl-89-067901,ku-nat-423-731,ja-prl-89-183601,im-prl-89-163602} and here we address the last by demonstrating the CNOT gate shown conceptually in Fig. \ref{schematic}(a)\cite{ra-pra-65-062324,ho-pra-66-024308}. Combined with quantum non-demolition (QND) measurement of the outputs, this gate is equivalent to the KLM CNOT: QND measurements can be made with additional single photon inputs, linear optics and number resolving single photon detection\cite{ko-pra-66-063814}. In Fig. \ref{schematic}(a) the control ($C$) and target ($T$) qubits act as their own ancilla: the gate operates correctly conditional on simultaneous detection of a single photon in each of the outputs, which is assumed in the following discussion. 

The gate works as follows\cite{ra-pra-65-062324}: The control and target qubits are each encoded by a single photon across two spatial modes---spatial encoding. As indicated in Fig. \ref{schematic}(b), any arbitrary superposition state is possible: $\alpha|0\rangle+\beta|1\rangle$, where $|0\rangle$ and $|1\rangle$ are the two spatial modes corresponding to the logical basis states. The coefficients are normalised complex amplitudes: $|\alpha|^2+|\beta|^2=1$. The two target modes are mixed and recombined on two 50\% reflective beam splitters ($\frac{1}{2}$BSs) to form an interferometer which also includes a 33\% reflective beam splitter ($\frac{1}{3}$BS) in each arm. This interferometer is balanced so that in the absence of a control photon, the target qubit exits in the same state that it entered. For the control in the state $|0\rangle$ this remains true because there is no interaction between the qubits. However, if the control is in the state $|1\rangle$ the control and target photons interfere non-classically at the central $\frac{1}{3}$BS due to path indistinguishability\cite{ho-prl-59-2044}. This two-photon quantum interference causes a $\pi$ phase shift in the upper arm of the target interferometer and results in the target state output being flipped: $\alpha|0\rangle+\beta|1\rangle\rightarrow\beta|0\rangle+\alpha|1\rangle$. The control qubit's logical value is unchanged. Because of the $\frac{1}{3}$BSs we do not always observe a single photon in each of the control and target outputs. However, when we do detect a single photon in each output (a coincidence count) which occurs with probability $P=\frac{1}{9}$, we know that the CNOT operation has been correctly realised. The most important feature of a CNOT gate is its quantum operation: with the control in a superposition and the target in a logical basis state, the output of the CNOT gate is an entangled state---the quintessential quantum mechanical state.

\begin{figure}[t!]
\begin{center}
\includegraphics*[width=7.5cm]{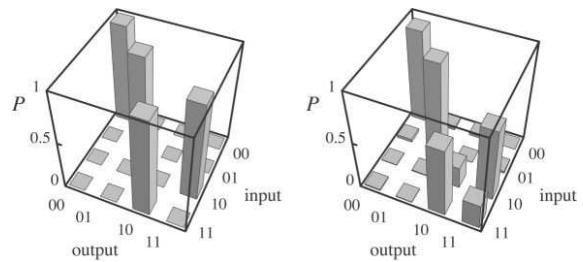}
\caption{Experimental demonstration of classical CNOT operation---operation in the logical basis. (a) Ideal logical basis operation of a CNOT gate. (b) Measured operation for the gate presented here.}
\label{truth}
\end{center}
\end{figure}

\begin{table}[b!]
  \centering 
  \caption{Experimentally determined probabilities for the logical basis operation, as plotted in Fig. \ref{truth}}\label{trutht}
\begin{center}
\begin{tabular}{|c|c|c|c|c|}
\hline
Input $|CT\rangle$  &$P_{|00\rangle}$  & $P_{|01\rangle}$ & $P_{|10\rangle}$ & $P_{|11\rangle}$ \\
\hline
\hline
$|00\rangle$   & 0.95(2)  & 0.023(3) & 0.024(3) & 0.0006(5) \\
\hline
$|01\rangle$   & 0.031(3) & 0.94(2) & 0.0019(8) & 0.022(3)\\
\hline
$|10\rangle$   & 0.005(1) & 0.011(2) & 0.23(9) & 0.75(2)\\
\hline
$|11\rangle$   & 0.011(2) & 0.0005(1) & 0.72(2) & 0.26(1)\\
\hline
\end{tabular}
\end{center}
\end{table}

It is most practical to prepare single photon qubits where the quantum information is encoded in the polarisation state $\alpha|H\rangle+\beta|V\rangle(\equiv\alpha|0\rangle+\beta|1\rangle)$---polarisation encoding, where $|H\rangle$ and $|V\rangle$ are the horizontal and vertical polarisation states, respectively. To convert to spatial encoding requires a polarising beam splitter (PBS) and half waveplate (HWP) as indicated in Fig. \ref{schematic}(b). To convert from polarisation, to spatial and back to polarisation encoding, while preserving the quantum information, requires that the phase relationship between the two basis components be preserved throughout: the path lengths must be sub-wavelength stable (interferometric stability). Therefore, the CNOT gate shown schematically in \ref{schematic}(a) requires two classical interferences, and one non-classical interference of the control and target photons, to be satisfied simultaneously---a significant challenge.

To meet these requirements we have designed the inherently stable interferometer shown in Fig. \ref{schematic}(c), where there is a one-to-one mapping from the conceptual schematic of Fig. \ref{schematic}(a). The target interferometer is realised by mixing and recombining the logical basis modes while the target qubit is polarisation encoded: This requires a HWP set to rotate the polarisation by 45$^\circ$ (ie. with its optic axis OA $=22.5^\circ$), which equally mixes the two polarisation modes (a Hadamard gate (H) in the language of quantum information: the CNOT gate can be thought of as a controlled-phase shift gate with a Hadamard gate at the input and output of the target). Transformation to spatial encoding occurs at a PBS where the two output modes are parallel but displaced. The operation of all three  $\frac{1}{3}$BSs is realised by a single ``$\frac{1}{3}$"HWP (OA=62.5$^\circ$):  The $C_1$ and $T_+$ components are orthogonally polarised and exit the first PBS in the same spatial mode. They are unequally mixed on the ``$\frac{1}{3}$"HWP and the required quantum interference is realised since the photons are only partly distinguishable after the second PBS. The OA setting is chosen to achieve this interference and also to swap the roles of H and V so that the control and target modes recombine correctly at the second PBS. In addition, the $C_0$ and $T_-$ modes are polarisation-rotated by the ``$\frac{1}{3}$"HWP such that, after the second PBS, $\frac{2}{3}$ of each component goes into the dumps, as required to balance the gate. The total 2-qubit output state of the gate is analysed via quantum state tomography.

To characterize the operation of this gate we first measured the output of the gate for each of the four possible logical basis input states: $|C\rangle |T\rangle =|CT\rangle =|00\rangle$, $|01\rangle$, $|10\rangle$ and $|11\rangle$. The correct behaviour is represented in Fig. \ref{truth}(a) and compared to that observed experimentally [Fig. \ref{truth}(b)]. The gate works very well for the $|C\rangle=|0\rangle$ ($94\pm2\%$ and $95\pm2\%$) inputs which reflects the fact that only a single classical interference is required for correct operation. For the $|C\rangle=|1\rangle$ inputs the gate works less well ($75\pm2\%$ and $72\pm2\%$), due to the added non-classical interference required (Table \ref{trutht}). The probability of getting the correct ouput averaged over all logical inputs is 84\%. This value compares favourably with that obtained from the data of Ref. \onlinecite{sc-nat-422-408}: 73.5\%. However, logical basis operation is purely classical and therefore demonstrates only part of the required action of a quantum CNOT gate. 

\begin{figure}[t!]
\begin{center}
\includegraphics*[width=6cm]{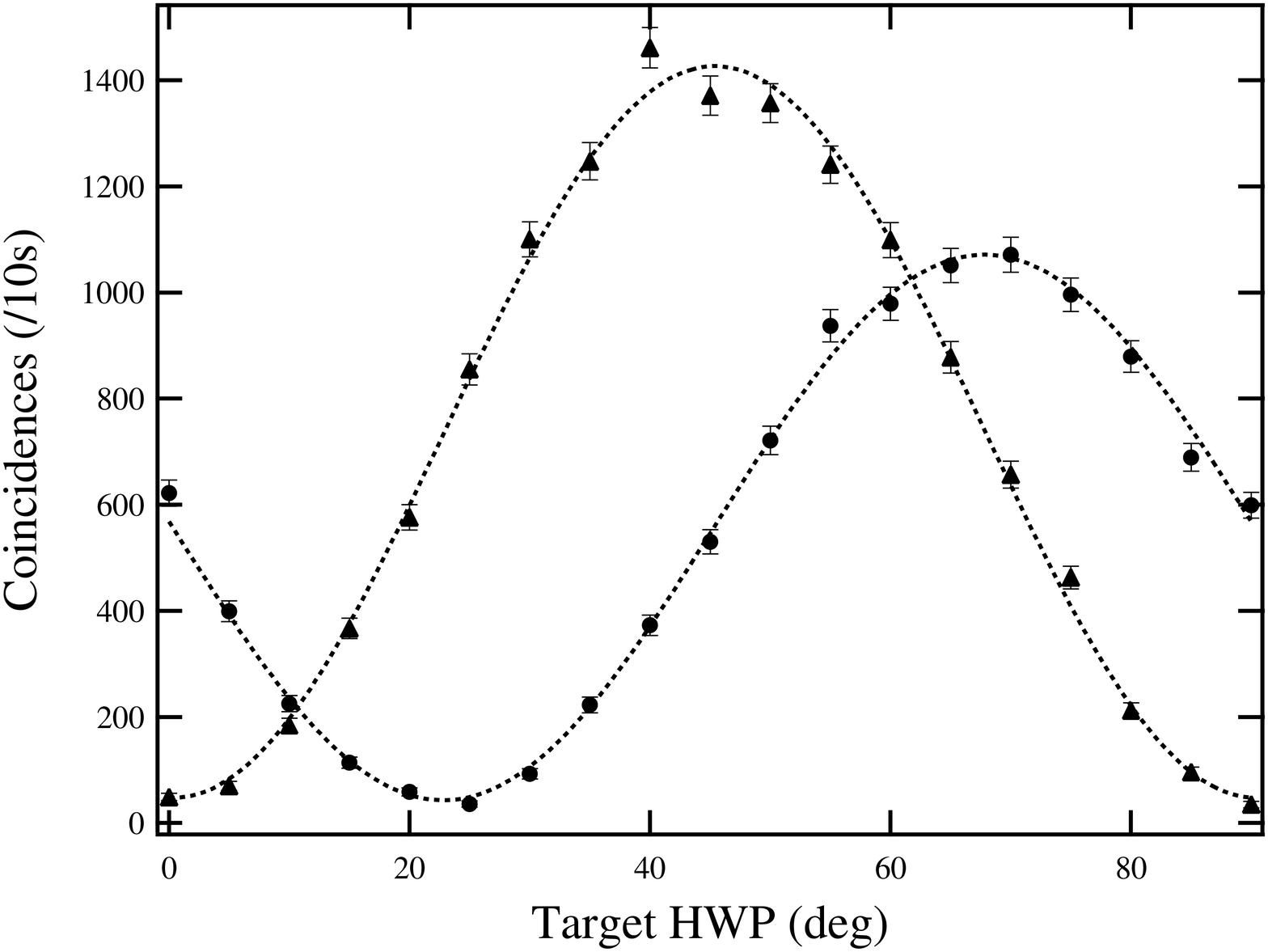}
\caption{Conditional coincidence fringes for non-orthogonal bases. The control analyser was set to pass $|0\rangle+|1\rangle$ (circles) and $|0\rangle$ (triangles) for the input state ($|0\rangle-|1\rangle)_C|1\rangle_T$. The error bars corresponding to HWP angle are too small to see on this plot. The fitted curves have a period which is fixed at 90$^\circ$ and the phase errors are $0.4\pm 0.2^\circ$ and $0.3\pm 0.2^\circ$, respectively.}
\label{fringe}
\end{center}
\end{figure}

\begin{figure}[t!]
\begin{center}
\includegraphics*[width=7.5cm]{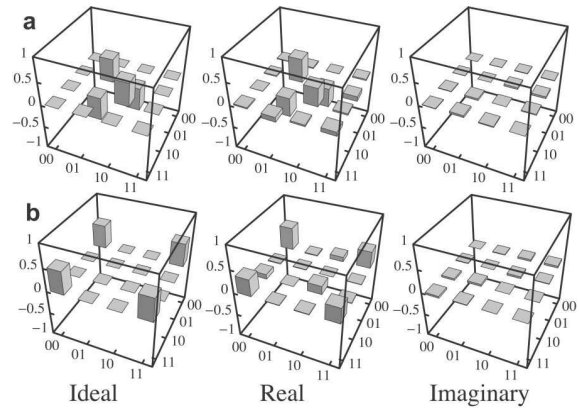}
\caption{Density matrices for highly entangled output states. (a) The real part of the density matrix for the maximally entangled Bell singlet state $|\Psi^-\rangle=|01\rangle-|10\rangle$ (the imaginary components are all zero), and the real and imaginary parts of the density matrix reconstructed from quantum process tomography for the input state $(|0\rangle-|1\rangle)_C|1\rangle_T$ . (b) The real part of the $|\Phi^+\rangle$ state, and the real and imaginary parts of the density matrix reconstructed from quantum process tomography for the input state $(|0\rangle+|1\rangle)_C|0\rangle_T$.}
\label{tomo}
\end{center}
\end{figure}

\begin{table}[b!]
  \centering 
  \caption{Characterisation of the four Bell states. The degree of entanglement of any state can be measured by calculating the Tangle $T=C^2$, where C is the concurrence\protect\cite{mu-jmo-48-1239,ja-pra-64-052312}. Similarly the degree of mixture can be measured by calculating the linear entropy\protect\cite{mu-jmo-48-1239,ja-pra-64-052312}. These values, along with overlap fidelities, for all four experimentally produced Bell states are given.}
\begin{center}
\begin{center}
\begin{tabular}{|c|c|c|c|}
\hline
State & Fidelity & Tangle & Linear Entropy\\
\hline
\hline
$\Psi^-=|01\rangle-|10\rangle$ & 0.87(8) & 0.65(6) & 0.27(6)\\
\hline
$\Psi^+=|01\rangle+|10\rangle$ & 0.75(9) & 0.55(7) & 0.31(5)\\
\hline
$\Phi^-=|00\rangle-|11\rangle$ & 0.76(9) & 0.46(5) & 0.45(3)\\
\hline
$\Phi^+=|00\rangle+|11\rangle$ & 0.77(9) & 0.49(9) & 0.45(5)\\
\hline
\end{tabular}
\end{center}
\end{center}
\label{bell}
\end{table}

The next step is to demonstrate that the gate is entangling --- it produces an entangled two qubit output state from a separable input --- which can be done by measuring conditional fringe visibilities\cite{ku-pra-64-023802}. Figure \ref{fringe} shows plots of coincident photon count rates as a function of the HWP setting in the target analyser. The two curves are for the control analyser HWP set to analyze  $|0\rangle$ and $|0\rangle+|1\rangle$, respectively. In both cases the input state is ($|0\rangle-|1\rangle)_C|1\rangle_T$, which ideally produces the maximally entangled Bell singlet state $|\Psi^-\rangle=|01\rangle-|10\rangle$. The visibilities ($\nu$=(max$-$min$)/($max$+$min)) for the fitted curves, for which the period is fixed at $90^\circ$, are $93.5\pm 2\%$ and $92\pm 3\%$ respectively. High visibility fringes in non-orthogonal bases like these are a classic signature of entanglement\cite{ku-pra-64-023802}, providing evidence for the quantum operation of the gate. However, they do not provide enough information to reconstruct the output state since the degree of mixture is unknown.

Complete state reconstruction is possible using quantum state tomography: a series of measurements on a large number of identically prepared copies of a quantum system allows accurate estimation of the quantum state of that system. In the case of two qubits, this requires 16 different joint measurements of the two qubit state\cite{ja-pra-64-052312} which can be used to reconstruct the density matrix which contains everything that can be known about the two qubit state. We propose that production and quantum state tomography of the four maximally entangled Bell states with high fidelity is an important functional demonstration for a CNOT gate. Figure \ref{tomo}(a) shows the density matrix of  $|\Psi^-\rangle$, and the real and imaginary parts of the reconstructed density matrix of the output state of our gate for the input $(|0\rangle-|1\rangle)_C|1\rangle_T$. The fidelity is $F_{\Psi^-}=\langle\Psi^-|\hat\rho|\Psi^-\rangle=0.87\pm0.08$. Also shown (b) are the density matrix for $|\Phi^+\rangle$ and the real and imaginary components of the reconstructed density matrix $\hat\rho$ for the input state $(|0\rangle+|1\rangle)_C|0\rangle_T$, where $F_{\Phi^+}=0.77\pm0.09$. The overlap fidelities for the other two Bell states (not plotted) are shown in Table \ref{bell}, along with measures of entanglement and mixture for all four Bell states. In all cases the experimentally measured output state is in the range where a Bell inequality can be violated\cite{mu-jmo-48-1239}.

All of the data shown here were taken over one day for a single operating condition of the gate and its performance was observed to be repeatable. This demonstrates two important points: the gate is very stable; and it does not require tuning for different input states. In additon, no correction has been made for accidental coincidence counts, which will introduce small errors in gate operation. By far the largest source of error in our gate is due to decoherence which arises from imperfect mode matching for the non-classical interference. This can be seen clearly in the logical basis operation shown in Fig. \ref{truth}: The gate works less well for states where the control is in the logical $|1\rangle$ state and the error terms are due to mode mismatch between the $C_1$ and $T_+$ modes, resulting in the target not being flipped as required. The errors in production of the $\Phi^+$ state are also due to mismatch of these modes: In the experimentally reconstructed density matrix [Fig. \ref{tomo}(b)] we can see that the error terms are residual components of the input state, with the original coherences preserved. The differences between the fidelities of the four Bell states --- which all require non-classical interference since $|C\rangle=|0\rangle\pm|1\rangle$ --- is understood to arise from small amounts of input beam steering introduced by the state preparation waveplates.

In summary, we have demonstrated a two photon CNOT gate operating via coincident photon detection. In the logical basis the gate operates with an average success of 84\%. Conditional fringe visibilities exceeding 90\% in non orthogonal bases indicate entanglement. Complete quantum state tomography confirms this, showing production of all four entangled Bell states with fidelities greater than 75\% --- an important functional demonstration of quantum CNOT operation. Note that these results go well beyond a recent report of an alternative optical CNOT gate\cite{pi-quant-ph-0303095} which shows logical basis operation, but only a single coincidence fringe with $\nu=61.5\pm7.4$\%, making the issue of entanglement ambiguous. The CNOT gate presented here combined with QND is equivalent to the KLM CNOT gate, and could be made scalable with the same teleportation protocol. It therefore provides the first experimental indication that all-optical quantum computing is possible. The next step will be to incorporate the gate demonstrated here in simple optical circuits to demonstrate simple algorithms and error correcting schemes\cite{do-quant-ph-0306081}.

\textbf{Acknowledgements}
We thank N. K. Langford for experimental work related to non-classical interference, T. B. Bell for work on the quantum state tomography system, and  P. T. Cochrane, J. L. Dodd, A. Gilchrist, P. G. Kwiat, G. J. Milburn, W. J. Munro and M. A. Nielsen for helpful discussions. This work was supported by the Australian government, the Australian Research Council, the US National Security Agency (NSA) and Advanced Research and Development Activity (ARDA) under Army Research Office (ARO) contract number DAAD 19-01-1-0651.

\end{document}